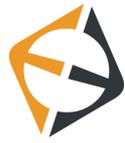

# Accelerated Option Pricing

# in

# *Multiple* Scenarios

04.07.2008

Stefan Dirnstorfer (stefan@thetaris.com)
Andreas J. Grau (grau@thetaris.com)

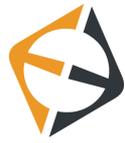

# Abstract

This paper covers a massive acceleration of Monte-Carlo based pricing method for financial products and financial derivatives. The method is applicable in risk management settings, where a financial product has to be priced under a number of potential future scenarios. Instead of starting a separate nested Monte Carlo simulation for each scenario under consideration, the new method covers the utilization of very few representative nested simulations and estimating the product prices at each scenario by a smoothing method based on the state-space. This smoothing technique can be e.g. non-parametric regression or kernel smoothing.

# Background

The currently most resource intensive computational task in risk management is pricing of financial products. Observable market parameters are simulated according to statistical properties based on historical observations. The simulated scenarios represent potential future values for which a financial institution has to prepare. Since each financial product in a large portfolio acts differently to changes in market parameters each product's price has to be evaluated under each scenario.

A number of regulatory provisions require financial institutions to perform a valuation of the portfolio under each potential future scenario. The accuracy and speed with which such evaluations can be performed is crucial to the companies financial success. As regulated by the Basel II directive each bank has to back its risky investments with risk-less assets such as government bonds. Since risky assets are generally expected to yield higher profits, banks are interested in measuring each asset's risk as accurately as possible. If banks fail to produce accurate risk figures, they are required to place large amounts of risk capital and consequently reducing the banks ability to invest large amounts into a particular asset. While Basel II is focused on the banking sector, similar regulations are planed for the insurance industry as proposed by Solvency II.

The current Basel II regulation requires a portfolio to be evaluated under a number potential future scenarios. For each scenario the development of all portfolio positions have to be determined for several time steps. It is currently best practice to randomly simulate future market parameters for each scenario and each time step. In a typical setting with 5000 paths and 250 time steps the total number of portfolio evaluations is given by 5000x250 = 1.25 million prices for each financial product in the portfolio. Since banks have portfolios with thousands of financial products, the computational challenge is enormous. There are a couple of financial products, such as European options or futures for which extremely fast algebraic evaluations can be performed. Other financial products can efficiently be priced by solving an associated partial differential equation. Other products, especially those that depend on a large number of traded instruments, e.g. basket options, can only be solved with a Monte Carlo approach. On prevailing computer hardware with a single processor one single Monte Carlo evaluation can take several minutes to compute and this is far to slow. The problem can only partially be mitigated through the use of multi-processor machines, since this process has to be performed for each individual financial asset in portfolio. The cost and size of today's computing hardware is already at a maximum in many banks, while risks of many exotic derivatives are still largely overestimated. For banks with a complete portfolio risk management system in place, the method of this paper will cut down the required computing time dramatically and thus reduces costs for maintenance and energy.

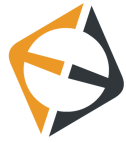

The method proposed in this document yields a massive reduction in the computational effort required to compute product prices in future scenarios as required e.g. for risk management. The method is an amendment to Monte Carlo based pricing. As such it can be applied to the same range of pricing problems, many of which can not be solved by other methods than Monte Carlo. However the new method provides an unprecedented speed-up. A computation that previously took months can be performed in a couple of seconds instead. Apart from being faster it is also more accurate than a couple of previously suggested solutions.

The speed-up of the method of this paper stems from the fact that a nested Monte Carlo simulation for each physical scenario for the risk assessment computes a lot of redundant information. Creating a pricing function of the financial products by smoothing of the simulated Monte Carlo values can be performed in a single step. Utilizing this pricing function instead of nested Monte Carlo simulations avoids the computation of the redundant information and thus accelerates the risk-management task.

Overall a bank will benefit from this method by the possibility to reduce its required risk capital. Any financial institution doing risk analysis will benefit from this method. It provides higher accuracy and reduces costs for performance computing hardware.

## Previous solutions

Depending on the instrument type and the time horizon of the risk estimation, the risk assessment of a product position in a portfolio can be conducted in several ways. In any case, risk is measured by a characteristic number, e.g. value at risk (VaR), conditional value at risk (CVaR) or standard deviation.

For a short time horizon, a risk estimate based on the sensitivities of the product with respect to the underlying ("Delta" and "Gamma") delivers fast and accurate results without the need of simulation. However, this sensitivity-based approach fails to estimate risks accurately when the remaining maturity time of the product is short or when the risk estimate for several weeks ahead has to be computed.

Estimating the market risk for long time spans, a simulation of the risk factors has to be conducted and at each time step of each scenario, the portfolio has to be evaluated. This is easy, if a fast pricing method for the specific instrument type exists, e.g. an analytic solution for the price. However, for many instrument types, only computational-expensive simulation methods exist, especially for Basket or path-dependent options. The cost of a simulation of the risk factors and a nested simulation for the product prices is in many realistic settings prohibitively high such that different solutions have been proposed to mitigate this problem. The important solutions are:

1. Usage of **variance reduction techniques** in the nested Monte Carlo simulation. Many variance reduction techniques have been proposed, e.g. control variables, low discrepancy sequences and importance sampling [Glasserman 2003]. But, the speed-up using the variance reduction techniques – typically between 2 and 10 - is by far not sufficient for a nested simulation estimating market risk.

2. **Portfolio compression**, which creates a new portfolio with the same risk properties as the considered portfolio but with fewer instruments [Dembo 1998]. This approach helps to some extend by reducing the number of instruments to price, but this technique is often not applicable to complex structured products.

3. Risk estimation by **combinatorial scenario simulation**, which effectively reduces the number of physical scenarios in settings with many risk factors. For each of the $s$ risk factor only a small number $n$ of physical scenarios is computed. Then the whole setting of the physical simulation is created by computing all possible combinatorial combinations of the riskfactors. Then, a Monte Carlo simulation along these precomputed risk factor realizations is performed to estimate the portfolio risk. This reduces the required number of option valuations considerably but shows slow convergence [Abken 2000].

4. **Importance sampling**, which computes scenario samples that are of particular importance for the risk measure to estimate and shifts the scenario weights such that the estimate of the risk measure is unbiased [Glassermann 2000]. This technique can improve the accuracy of risk estimates considerably but the speed-up is often still not sufficient for nested Monte Carlo simulations.

5. **Usage of few paths** is another method for mitigating the computational burden. It turns out that if at each of the physical paths, a nested simulation with e.g. 100 paths is conduced (where e.g. 10.000 would be needed for a precise option price estimate) can already lead to sufficiently precise risk estimates. The reason for this is that the errors in the option price estimations mutually annihilate almost completely and the few paths are sufficient for estimates of the risk measure. However, the resulting risk-measure estimate is biased and has to be corrected in a post process for precise estimates. [Gordy and Juneja 2008]

# The new method

The task the method of this paper performs efficiently is defined as follows: Consider that a financial product shall be priced under many different future market scenarios and at different time-steps. Instead of starting a separate nested Monte Carlo simulation for each scenario under consideration, the new method covers the utilization of one or more well selected nested simulations. Based on the simulations, a function will be derived which maps the state of all relevant variables onto a price estimate for the financial product. Such a function can be built from e.g. non-parametric regression or kernel smoothing. In realistic settings, the overhead of building this function is small and substantial performance improvements are gained: Simply evaluating the estimated function is quicker than evaluation by nested Monte Carlo simulation, even if the nested simulation uses only very few paths.

## Overview

The algorithm for the pricing of a financial product in multiple scenarios comprises 7 separate steps:

1. Importing scenarios $P : \mathbb{T}_p \times \Omega_p \to \mathbb{R}^s$ at which the product prices shall be computed. Each element of $\Omega_p$ is associated with a scenario for $s$ risk-factors drawn at each time step in $\mathbb{T}_p$.
2. Generation of scenarios $Q : \mathbb{T}_q \times \Omega_q \to \mathbb{R}^s$ for the product price estimation. $Q$ is computed by sampling of a stochastic process. Each element of $\Omega_q$ is associated with a scenario for $s$ risk-factors drawn at each time step in $\mathbb{T}_q$.
3. Compute path-dependent product specific variables $A_p : \mathbb{T}_p \times \Omega_p \to \mathbb{R}^{s_a}$ corresponding to

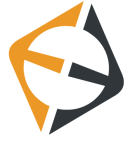

the scenarios in step 1. These path-dependent variables include for example fixings or exercises performed by the product's issuer or the holder.

4. For each scenario of step 2: Computation of product specific variables $A_q : \mathbb{T}_q \times \Omega_q \to \mathbb{R}^{s_a}$ corresponding to the scenario. These path-dependent variables include fixings and optimal exercises.
5. Computation of the product's remaining discounted cash-flows $V_q : \mathbb{T}_q \times \Omega_q$. At each scenario and each time-step in the scenarios of step 2 all remaining cash-flows are scenario-wise discounted and summed.
6. For each scenario of step 1: Computation of a price estimate function $F : \mathbb{T}_p \times \mathbb{R}^s \times \mathbb{R}^{s_a} \to \mathbb{R}$. $F$ is obtained by a smoothing procedure on the scenarios of step 2 and the path-dependent variables of step 4.
7. Computation of the product prices for each scenario $\omega_p \in \Omega_p$ and each time step $t_p \in \mathbb{T}_p$ from step 1. This evaluation is performed efficiently by
$$V_p(t_p, \omega_p) = F(t_p, P(t_p, \omega_p), A_p(t_p, \omega_p)).$$

## Details of the Steps

In **step 1**, the scenarios consist of a realization of values for each risk factor which has to be taken into account. Typical risk factors for a structured financial product are: prices of underlyings, implied volatilities and long-term as well as short term interest-rates. In the following, we will refer to the scenarios from step 1 as physical scenarios and all associated variables are denoted by an index $p$.

The origin of the scenarios in step 1 can be manifold: historical simulation, shifting of current risk factor values and Monte Carlo simulation are possible choices. The scenarios can consist of a single time step or multiple time steps. The particular choice depends on the specific result one expects from the analysis. A multi-time step Monte Carlo simulation might be useful for the computation of risk-measures such as Value at Risk while a single time step which a shift of the risk factors is useful for stress testing and estimating the risk contribution of single instruments.

The physical scenarios are denoted by $P : \mathbb{T}_p \times \Omega_p \to \mathbb{R}^s$, whereas $\Omega_p = \{1, \dots, n_p\}$ is a numbering for the scenarios and $\mathbb{T}_p = \{t_p^0, \dots, t_p^{T_p}\}$ is the set of time steps. At each scenario and each time step, an $s$-tuple of risk-factors is given.

The scenarios of **step 2** are used for the product valuation itself and it is useful to generate so called risk-neutral scenarios for this task as defined by the option pricing theory. All associated variables are denoted by an index $q$. Examples for such scenarios are e.g. geometric Brownian motion where the drift is set to the risk-free rate of interest and constant volatility as well as geometric Brownian motion with Heston volatility [Heston 1993].

The scenarios of step 2 are denoted by $Q : \mathbb{T}_q \times \Omega_q \to \mathbb{R}^s$, whereas $\Omega_q = \{1, \dots, n_q\}$ is a numbering for the scenarios and $\mathbb{T}_q = \{t_q^0, \dots, t_q^{T_q}\}$ is the set of time steps. At each scenario and each time step, an $s$-tuple of risk-factors is sampled from a stochastic model. Additionally, there is a mapping $I : \Omega_q \to \mathbb{T}_q$. For each scenario $\omega_q \in \Omega_q$, the scenario path $Q(t, \omega_q), t \in \mathbb{T}_q$ is called active for $t \geq I(\omega_q)$.

The set $\mathbb{T}_q$ contains all relevant time steps (fixings) for the evaluation of the financial product. Furthermore, the algorithm works well when relevant physical scenario time steps are contained as well, i.e. $\mathbb{T}_q \supset \{t_p^0, t_p^1, \ldots, t_p^k\}$, $t_p^k \leq t_q^{T_q}$, where $t_q^{T_q}$ is the maturity time of the financial product.

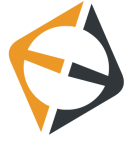

For step 2, an implementation to create the simulations with one or more of the following properties can be beneficial:

  a. the scenario paths start at the same time and the same value as the physical scenarios, i.e. $I(\omega_q) = t_p^0$ and $P(t_p^0, \omega_p) = Q(t_p^0, \omega_q)\ \forall \omega_p \in \Omega_p,\ \forall \omega_q \in \Omega_q$,
  b. the scenario paths start at the same time and similar value as the physical scenarios, i.e. $I(\omega_q) = t_p^0$ and $P(t_p^0, \omega_p) \approx Q(t_p^0, \omega_q)\ \forall \omega_p \in \Omega_p,\ \forall \omega_q \in \Omega_q$,
  c. each scenario path $\omega_q \in \Omega_q$ forks a physical scenario at some time $t_\omega \in \mathbb{T}_q \cap \mathbb{T}_p$, i.e. $I(\omega_q) = t_\omega$ and $\exists \omega_p \in \Omega_p : P(t_\omega, \omega_p) = Q(t_\omega, \omega_q)$
  d. each scenario path $\omega_q \in \Omega_q$ forks at some time $t_\omega \in \mathbb{T}_q \cap \mathbb{T}_p$ in the proximity of a physical scenario, i.e. $I(\omega_q) = t_\omega$ and $\exists \omega_p \in \Omega_p : P(t_\omega, \omega_p) \approx Q(t_\omega, \omega_q)$.

In **step 3**, for time step $t_p \in \mathbb{T}_p$ and each scenario $\omega_p \in \Omega_p$ the path dependent values $A_p : \mathbb{T}_p \times \Omega_p \to \mathbb{R}^{s_a}$ are computed. These $s_a$-tuples together with the current risk factor values $P(t_p, \omega_p)$ must be sufficient to price the financial product at time step $t_p$. The values at time $t_p^0$ are given by an initialization function $f_0$, i.e. $A_p(t_p^0, \omega_p) = f_0(t_p^0, P(t_p^0, \omega_p))$. These values are part of the product's structural features and could be imported from a database. The successive values at times $t_p^i > t_p^0$ are computed from an update formula $f_p$, i.e. $A_p(t_p^i, \omega_p) = f_p(t_p^i, A_p(t_p^{i-1}, \omega_p), P(t_p^i, \omega_p))$ for $i = 1, \ldots, T_p$.

Examples for path dependent variables $A_p$ are
  - information about knock-out of Barrier options,
  - the current average of Asian options,
  - exercise, conversion and calls of the financial product based on e.g. the investor's utility,
  - measurable characteristic values w.r.t. the stochastic model in step 2,
  - portfolio weights of dynamic strategies, e.g. Simulation-Based Hedging (Grau 2008),
  - previous values of risk factors.

In **step 4**, similar to step 3, the path-dependent variables $A_q : \mathbb{T}_q \times \Omega_q \to \mathbb{R}^{s_a}$ for time step $t_q \in \mathbb{T}_q$ and each scenario $\omega_q \in \Omega_q$. These $s_a$-tuples together with the current risk factor values $P(t_q, \omega_q)$ must be sufficient to price the financial product at time step $t_q$. The values at time $I(\omega_q)$ are initialized with appropriate values.

For an implementation to compute the initial path-dependent variables $A_q$ it can be beneficial to use one of the following methods:

  a. If there exists at least one physical path that matches a scenario path at its first active time step then we can use a physical path-dependent state as initial state, i.e,
  $\exists \omega_p \in \Omega_p : P(t_\omega, \omega_p) = Q(t_\omega, \omega_q),\ I(\omega_q) = t_\omega$ then use $A_q(t_\omega, \omega_q) = A_p(t_\omega, \omega_p)$.
  b. Alternatively, we can choose a physical path $\omega_p$ which is similar to the scenario paths $\omega_q$ at time $t_\omega = I(\omega_q)$. Then, we initialize the path-dependent state $A_q(t_\omega, \omega_q)$ to be equal or similar to $A_p(t_\omega, \omega_p)$, i.e. $\exists \omega_p \in \Omega_p : P(t_\omega, \omega_p) \approx Q(t_\omega, \omega_q),\ I(\omega_q) = t_\omega$ then use $A_q(t_\omega, \omega_q) \approx A_p(t_\omega, \omega_p)$ where $A_q$ is an artificial realization of the path-dependent variables. Note that the new values should be consistent with the structure of the financial product and possible path histories.

c. For each scenario path $\omega_q$, we initialize the path-dependent variables from a synthetic path $R_\omega : \mathbb{T}_r \to \mathbb{R}^s$ with $\{t_p^0, t_\omega\} \subset \mathbb{T}_r$. The synthetic path must have the same value as the scenario path at time $t_\omega$, i.e. $R_\omega(t_\omega) = Q(t_\omega, \omega_q)$. Equivalent to the update formula in step 3, we compute $A_q(t_\omega, \omega_q) = A_r(t_\omega)$ through an iterative process.

In **step 5**, the product's remaining cash-flows are discounted to a cash value $V_q : \mathbb{T}_q \times \Omega_q$. Consider a discount factor $d : \mathbb{T}_p \times \mathbb{T}_q \times \Omega_q \to \mathbb{R}^+$. For each scenario $\omega_q \in \Omega_q$ returns the function $d(t_p, t_q, \omega_q)$ the discount factor from time $t_q$ to time $t_p$. This function is constructed knowing the full history of the path $\omega_q$.

In each scenario the product-specific cash flows are given by $C : \Omega_q \times \mathbb{T}_q \to \mathbb{R}$, i.e. in scenario $\omega_q$ pays $C(\omega_q, t_q)$ at time $t_q$. The cumulated and discounted remaining cash flows V are computed by

$$V_q(t_q, \omega_q) = \sum_{\substack{t \in \mathbb{T}_q \\ t \geq t_q}} d\left(t_p^0, t, \omega_q\right) C(t, \omega_q).$$

The crucial part of the algorithm is **step 6** where the product prices in each physical scenario and each physical time-step are computed using the scenario paths from step 2. Consider subsets $\widetilde{\mathbb{T}} \subset \mathbb{T}_q$ and $\widetilde{\Omega} \subset \Omega_q$. The set $M(\widetilde{\mathbb{T}}, \widetilde{\Omega})$ is defined as a set of (X, Y)-pairs that can be used for smoothing algorithms,

$$\begin{aligned} M(\widetilde{\mathbb{T}}, \widetilde{\Omega}) \;=\; \{(X, Y) \;:\; X \;&=\; (t, Q(t,\omega), A_q(t,\omega)) \;\in \mathbb{R}^{1+s+s_a}, \\ Y \;&=\; V_q(t,\omega), \\ t \;&\in\; \widetilde{\mathbb{T}},\; t \leq I(\omega), \\ \omega \;&\in\; \widetilde{\Omega}\}. \end{aligned}$$

The operator $\Psi$ computes a smoothing on a set of (X, Y)-pairs which results in a function mapping risk factor tuples and path-dependent state tuples onto product prices, i.e

$$\Psi : \left(\mathbb{R}^{1+s+s_a} \times \mathbb{R}\right)^n \to \left(\mathbb{R}^{1+s+s_a} \to \mathbb{R}\right).$$

It is useful that $\Psi$ creates an estimator for the conditional expected values $E(X|Y)$. Useful smoothing algorithms for $\Psi$ are:

a. Non-parametric regression sets the result function as a linear combination of basis functions $b_i$, i.e. $\Psi(M) = \sum_{i=1}^{n_b} c_i b_i$. The coefficients $c_i$ are determined by minimizing the quadratic error

$$(c_1, \ldots, c_{n_b}) = \operatorname{argmin} \sum_{(X,Y) \in M} \left(\Psi(M)(X) - Y\right)^2.$$

b. Kernel smoothing is defined by a sum of weighted $Y$ values, i.e. $\Psi(M)(x) = \frac{1}{|M|} \sum_i w_i(x) Y_i$, with a weight function $w_i(x)$ constructed from the location of the $X$ values.

Further information about the smoothing algorithms mentioned here and further smoothing algorithms can be found at Härdle 2001. An interesting approach to non-parametric regression is presented by Garcke et al 2001. Sometimes it is useful to select a subset of $M$ before performing one of the above smoothing algorithms. Furthermore, it also can be useful to use semi-parametric regression, thin-plate splines or b-spline basis functions.

The function $F: \mathbb{T}_p \times \mathbb{R}^s \times \mathbb{R}^{s_a} \to \mathbb{R}$ allows the efficient evaluation of prices in all time-steps and all physical scenarios. It can be constructed in one of the following ways:

a. The smoothing is done on all data at once, i.e. $F(t, Q, A) = \Psi\left(M(\mathbb{T}_q, \Omega_q)\right)(t, Q, A)$.

b. The smoothing is done on each time step individually, i.e. $F(t, Q, A) = \Psi\left(M(\{t\}, \Omega_q)\right)(t, Q, A)$. With only a single time step per smoothing, regression methods benefit from decreasing the dimensionality by 1.

c. Other partitions of $\mathbb{T}_p$ and $\Omega_p$ might be useful to cut the large smoothing problem into a set of smaller smoothing problem.

Finally, the product price is computed in **Step 7.** For each scenario $\omega_p \in \Omega_p$ and each time step $t_p \in \mathbb{T}_p$, the evaluation is performed efficiently as $V_p(t_p, \omega_p) = F(t_p, P(t_p, \omega_p), A_p(t_p, \omega_p))$. The price estimates $V_p$ are computed within the stochastic model generating the scenario paths in step 2. Hence, this algorithm is an efficient way to compute product prices in physical scenarios based on an arbitrary stochastic model.

It can be useful that the scenario paths $Q$ and the associated path-dependent variables $A_q$ are made persistent such that later computations of the smoothing function can be performed efficiently. Another possibility for an improvement is make the smoothing function $F$ persistent itself such that later computations of product prices for new risk factor tuples $P$ can be performed efficiently. Then, is can be useful to refine the smoothing function $F$ iteratively by computing additional scenario paths on demand, based on an error estimate for the price generated at the new risk factor tuples.

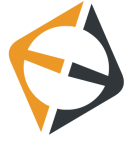

# Numerical examples

The following section describes a detailed example of the computation of financial product prices by Monte Carlo simulation in several physical scenarios using the method of this paper. In order to present a concise and reproduce able example, we restrict the example to 3 physical and 5 risk-neutral scenarios. It is easy to extend this small example to a realistic setting by adding more scenarios and additional risk-factors.

## Example

Consider a European call option with a strike price of 100 and a maturity time of 3 years. We use the physical scenarios $\Omega_p = \{1, 2, 3\}$ and the time-steps $\mathbb{T}_p = \{t_0, t_1, t_2\}$. Consider further the possible values of the physical scenarios $P$ for a stock price, which serves as the underlying of the European option:

| $P(t_p, \omega_p)$ | | $t_p$ | | |
|---|---|---|---|---|
| | | $t_0$ | $t_1$ | $t_2$ |
| $\omega_p$ | 1 | 100 | 110 | 120 |
| | 2 | 100 | 100 | 100 |
| | 3 | 100 | 90 | 80 |

In each of these scenarios at each time-step the European option value shall be estimated by Monte Carlo simulation. The 6 option prices for time steps $t_1$ and $t_2$ shall be computed as fast and accurate as possible. Prior art would perform 6 completely separate pricing procedures. The method of this paper only requires a single scenario set $Q$ of risk-neutral scenario paths $\Omega_q = \{1, 2, 3, 4, 5\}$ at time steps $\mathbb{T}_q = \{t_0, t_1, t_2, t_3\}$ starting at time $t_0$, i.e. $I(1) = I(2) = I(3) = I(4) = I(5) = t_0$:

| $Q(t_q, \omega_q)$ | | $t_q$ | | | |
|---|---|---|---|---|---|
| | | $t_0$ | $t_1$ | $t_2$ | $t_3$ |
| $\omega_q$ | 1 | 100.0000 | 211.7568 | 214.8651 | 106.2542 |
| | 2 | 100.0000 | 112.9350 | 70.6952 | 70.8322 |
| | 3 | 100.0000 | 154.1112 | 193.8189 | 221.6990 |
| | 4 | 100.0000 | 90.2616 | 155.3396 | 121.7245 |
| | 5 | 100.0000 | 174.4274 | 199.2726 | 258.4810 |

These risk-neutral scenarios are created using a stochastic model with geometric Brownian motion for $Q$ but other (risk-neutral) simulations are suitable, too. Now, we compute the payoff of the option value at time $t_3$, $C(t_3, \omega_q) = \max(Q(t_3, \omega_q) - 100, 0)$, which is equal to $V_q(t, \omega_q)$ for all $t \in \mathbb{T}_q$ because there is only a single cash-flow at maturity time and the risk-free rate of interest is zero ( $d(t_0, t, \omega_q) = 1$). Note that the option has no path-dependency, thus $A$ is empty and $s_a = 0$. The values are:

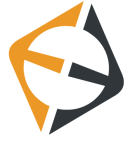

| $\omega_q$ | $C(t_3, \omega_q)$ | $V_q(t, \omega_q)$ |
|---|---|---|
| 1 | 6.2542 | 6.2542 |
| 2 | 0.0000 | 0.0000 |
| 3 | 121.6990 | 121.6990 |
| 4 | 21.7245 | 21.7245 |
| 5 | 158.4810 | 158.4810 |

In order to obtain estimates of the option prices at time $t_2$, we create the set $M(\{t_1\}, \Omega_q)$:

$$
\begin{array}{cc}
M(\{t_1\}, \Omega_q) \, X = (t, Q, A) & Y \\
(t_1, 211.7568) & 6.2542 \\
(t_1, 112.9350) & 0.0000 \\
(t_1, 154.1112) & 121.6990 \\
(t_1, 90.2616) & 21.7245 \\
(t_1, 174.4274) & 158.4810
\end{array}
$$

Next, the smoothing operation $\Psi$ has to be applied to the data set $M$. Since option pricing is performed in a Black-Scholes setting, option prices are given by conditional expected values $E(Y|X)$. Thus, estimates for the expected values are also estimates for the option price. Here, we use a simple nonparametric regression in $X_2$. $X_1$ is constant and will not be considered. The smoothing functions is

$$\Psi(M) = c_1 + c_2 \cdot X_2 + c_3 \cdot (X_2)^2$$

where $c_1$, $c_2$ and $c_3$ are coefficients of polynomial basis functions. In a realistic setting, other smoothing methods or other basis functions are useful, too. Now, we compute the coefficients as a solution to the minimization of

$$(c_1, c_2, c_3) = \mathrm{argmin} \left( \sum_{\omega=1}^{5} (Y_\omega - \Psi(M)(X_\omega))^2 \right)$$

for all risk-neutral scenarios $\omega = 1 \ldots 5$. This least-squares minimization is a standard problem and can be solved by normal equations using the matrix

$$B_q = \begin{pmatrix} 1 & X_{1,2} & (X_{1,2})^2 \\ \vdots & \vdots & \vdots \\ 1 & X_{5,2} & (X_{5,2})^2 \end{pmatrix},$$

$$B_q(t_1) = \begin{pmatrix} 1 & 211.7568 & 44840.9334 \\ 1 & 112.9350 & 12754.3143 \\ 1 & 154.1112 & 23750.2631 \\ 1 & 90.2616 & 8147.1538 \\ 1 & 174.4274 & 30424.9189 \end{pmatrix}.$$

Now, the coefficients $c_1$, $c_2$ and $c_3$ can be obtained by

$$c = (B_q^T B_q)^{-1} \cdot B_q^T V$$

which leads to

$$\begin{aligned} c_1 &= -651.7604 \\ c_2 &= 9.9033 \\ c_3 &= -0.0317 \end{aligned}$$

In order to obtain the approximations for the physical asset paths, we need to set-up the corresponding matrix with one row for each physical scenario

$$B_p = \begin{pmatrix} 1 & P(t_1, 1) & (P(t_1, 1))^2 \\ 1 & P(t_1, 2) & (P(t_1, 2))^2 \\ 1 & P(t_1, 3) & (P(t_1, 3))^2 \end{pmatrix}.$$

This leads to

$$B_p(t_1) = \begin{pmatrix} 1 & 110 & 12100 \\ 1 & 100 & 10000 \\ 1 & 90 & 8100 \end{pmatrix}$$

and the approximations of the smoothing function $F = \Psi(M)$ are obtained for the first time step, by

$$\begin{pmatrix} F(t_1, P(t_1, 1)) \\ F(t_1, P(t_1, 2)) \\ F(t_1, P(t_1, 3)) \end{pmatrix} = B_p \cdot c.$$

The result is the option price estimation $V_p(t_1, \omega_p)$ for each scenario $\omega_p \in \Omega_p$,

| $\omega_p$ | $V_p(t_1, \omega_p)$ |
|---|---|
| 1 | 54.57 |
| 2 | 22.01 |
| 3 | - 16.87 |

This is a very efficient way of estimating option prices. Note that 5 scenarios and 3 basis functions are not sufficient for precise estimates. This simplified example results in a negative price estimate for physical scenario 3.

Now, we can use the same scenarios to obtain the prices at time $t_2$ of the physical scenarios. In order to obtain estimates of the option prices at time $t_2$, we create the set $M(\{t_2\}, \Omega_q)$:

$$
\begin{array}{ccc}
M(\{t_2\}, \Omega_q) & X = (t, Q, A) & Y \\
& (t_2, 214.8651) & 6.2542 \\
& (t_2, 70.6952) & 0.0000 \\
& (t_2, 193.8189) & 121.6990 \\
& (t_2, 155.3396) & 21.7245 \\
& (t_2, 199.2726) & 158.4810
\end{array}
$$

Again, we construct a matrix $B_q$ for the regression

$$
B_q(t_2) = \begin{pmatrix}
1 & 214.8651 & 46167.0019 \\
1 & 70.6952 & 4997.8131 \\
1 & 193.8189 & 37565.7663 \\
1 & 155.3396 & 24130.3830 \\
1 & 199.2726 & 39709.5530
\end{pmatrix}
$$

and solving

$$
(c_1, c_2, c_3) = \text{argmin}\left(\sum_{\omega=1}^{5}(Y_\omega - \Psi(M)(X_\omega))^2\right)
$$

with the scenarios at time $t_2$ leads to

$$
\begin{aligned}
c_1 &= -\ 137.9136 \\
c_2 &= \ \ \ 2.2651 \\
c_3 &= -\ 0.0058
\end{aligned}
$$

With the corresponding matrix for the physical scenarios

$$
B_p(t_2) = \begin{pmatrix}
1 & 120 & 14400 \\
1 & 100 & 10000 \\
1 & 80 & 6400
\end{pmatrix}
$$

the option price estimates $B_p \cdot c$ evaluate to

| $\omega_p$ | $V_p(t_2.\omega_p)$ |
|---|---|
| 1 | 49.77 |
| 2 | 30.17 |
| 3 | 5.90 |

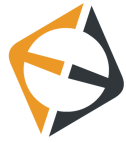

Concluding this first example, we obtain for each physical scenario and each time-step, an option price

| $\omega_p$ | $P(t_1,\omega_p)$ | $V_p(t_1,\omega_p)$ | $P(t_2,\omega_p)$ | $V_p(t_2.\omega_p)$ |
|---|---|---|---|---|
| 1 | 110 | 54.57 | 120 | 49.77 |
| 2 | 100 | 22.01 | 100 | 30.17 |
| 3 | 90 | -16.87 | 80 | 5.90 |

## Extension 1

The example can be extended in several ways. First of all, we can change the example to utilizing risk-neutral scenarios starting at different initial values, i. e.

| $Q(t_q,\omega_q)$ | | $t_q$ | | | |
|---|---|---|---|---|---|
| | | $t_0$ | $t_1$ | $t_2$ | $t_3$ |
| $\omega_q$ | 1 | **80.0000** | 64.1116 | 115.4375 | 105.6911 |
| | 2 | **90.0000** | 41.3639 | 72.6489 | 100.4893 |
| | 3 | **100.0000** | 105.1411 | 103.5702 | 81.8529 |
| | 4 | **110.0000** | 122.1953 | 137.8884 | 316.2593 |
| | 5 | **120.0000** | 83.2175 | 87.7915 | 84.1920 |

These scenarios can be used in the exact same way as before. Using the method with such a risk-neutral scenario set can lead to considerably higher precision of the option prices in extreme physical scenarios.

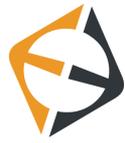

## Extension 2

Another extension of the example again considers the risk-neutral scenarios. In some settings it is beneficial to create additional scenarios in the risk-neutral setting at the exact value of the physical scenario, i. e.

| $Q(t_q, \omega_q)$ | | $t_q$ | | | | |
|---|---|---|---|---|---|---|
| | | $t_0$ | $t_1$ | $t_2$ | $t_3$ | $I(\omega_q)$ |
| $\omega_q$ | 1 | 100.0000 | 211.7568 | 214.8651 | 106.2542 | $t_0$ |
| | 2 | 100.0000 | 112.9350 | 70.6952 | 70.8322 | $t_0$ |
| | 3 | 100.0000 | 154.1112 | 193.8189 | 221.6990 | $t_0$ |
| | 4 | 100.0000 | 90.2616 | 155.3396 | 121.7245 | $t_0$ |
| | 5 | 100.0000 | 174.4274 | 199.2726 | 258.4810 | $t_0$ |
| | 6 | | **110.0000** | 139.2342 | 149.1234 | $t_1$ |
| | 7 | | **100.0000** | 78.9872 | 90.2324 | $t_1$ |
| | 8 | | **90.0000** | 98.9079 | 78.2347 | $t_1$ |
| | 9 | | | **120.0000** | 98.8968 | $t_2$ |
| | 10 | | | **100.0000** | 76.2563 | $t_2$ |
| | 11 | | | **80.0000** | 87.2342 | $t_2$ |

The scenarios 6 to 11 are added to the scenario set $\widetilde{\Omega}$ in order to fit the physical scenarios 1 to 3. Similar to the first extension of this example, the utilization of the method of this paper can ensure higher accuracy for extreme scenarios. Note that scenarios 9-11 are not utilized for the pricing at time-step $t_1$.

# Extension 3

A third extension to the example is required for the pricing of a path dependent option. Consider an Asian option, which has a payoff depending on the average asset price until maturity time of the option. This means that the current average $A_p$ must be computed for the physical as well as $A_q$ the risk-neutral simulations. For the physical scenarios $A_p$ is given by

| $\omega_p$ | $A_p(t_0,\omega_p)$ | $A_p(t_1,\omega_p)$ | $A_p(t_2,\omega_p)$ |
|---|---|---|---|
| 1 | 100 | 105 | 110 |
| 2 | 100 | 100 | 100 |
| 3 | 100 | 95 | 90 |

and for the risk-neutral scenarios $A_q$ is given by

| $\omega_q$ | $A_q(t_0,\omega_q)$ | $A_q(t_0,\omega_q)$ | $A_q(t_0,\omega_q)$ | $A_q(t_0,\omega_q)$ |
|---|---|---|---|---|
| 1 | 100.0000 | 155.8784 | 175.5406 | 158.2190 |
| 2 | 100.0000 | 106.4675 | 94.5434 | 88.6156 |
| 3 | 100.0000 | 127.0556 | 149.3100 | 167.4073 |
| 4 | 100.0000 | 95.1308 | 115.2004 | 116.8314 |
| 5 | 100.0000 | 137.2137 | 157.9000 | 183.0452 |
| 6 | 100.0000 | 105.0000 | 116.4114 | 124.5894 |
| 7 | 100.0000 | 100.0000 | 92.9957 | 92.3049 |
| 8 | 100.0000 | 95.0000 | 96.3026 | 91.7857 |
| 9 | 100.0000 | 105.0000 | 110.0000 | 107.2242 |
| 10 | 100.0000 | 100.0000 | 100.0000 | 94.0641 |
| 11 | 100.0000 | 95.0000 | 90.0000 | 89.3085 |

Note that the values for $A_q$ at time $t_2$ in scenarios 9 to 11 can be obtained directly from the physical scenarios. This ensures that the added scenarios are consistent with the other scenarios and that they are still increasing the numerical accuracy of the prices in extreme scenarios.

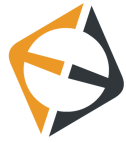

Computing the payoff $V(t, \omega_q) = C(t_3, \omega_q) = \max(A_q(t_3, \omega_q) - 100, 0), t = t_1, t_2, t_3$

| $\omega_q$ | $V_q(t, \omega_q)$ |
|---|---|
| 1 | 58.2190 |
| 2 | 0 |
| 3 | 67.4073 |
| 4 | 16.8314 |
| 5 | 83.0452 |
| 6 | 24.5894 |
| 7 | 0 |
| 8 | 0 |
| 9 | 7.2242 |
| 10 | 0 |
| 11 | 0 |

In oder to create a simple regression method we set up our data set as $M = (X, Y)_{\omega_q \in \Omega_q}$, whereas $X = (t_1, Q(t_1, \omega_q), A_q(t_1, \omega_q))$ and $Y = V_q(t_1, \omega_q)$. We compute the regression

$$\Psi(M) = c_1 + c_2 X_2 + c_3 (X_2)^2 + c_4 X_3 + c_5 (X_3)^2 + c_6 X_2 X_3$$

in order to estimate $E(Y|X)$, i.e. the option prices in each scenario.

At time step $t_1$, this leads to the coefficients

$$\begin{aligned} c_1 &= 0 \\ c_2 &= 0 \\ c_3 &= -0.0510 \\ c_4 &= 0 \\ c_5 &= -0.0739 \\ c_6 &= 0.1257 \end{aligned}$$

and thus to the option price estimates of

| $\omega_p$ | $V_p(t_1, \omega_p)$ |
|---|---|
| 1 | 20.28 |
| 2 | 8.24 |
| 3 | -5.13 |

The option prices $V_p(t_2, \omega_p)$ can be obtained correspondingly.

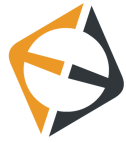

## Conclusion

This paper outlines a technique for estimating option prices for many scenarios by an efficient function smoothing method on risk-neutral Monte Carlo simulations. The resulting technique has much lower computational complexity by avoiding nested Monte-Carlo simulations. Numerical examples show that the new technique can be applied in realistic settings and that the gain in speed is substantial. Compared with previous techniques, this new technique can be applied in almost any setting, where option pricing by Monte Carlo simulation is applicable. The previously proposed solutions like nested Monte Carlo simulation with few paths or valuation by solving partial differential solutions (PDEs) are applicable only in much more restrictive settings.